# Nonlinear generalized functions and nonlinear numerical simulations in fluid and solid continuum mechanics

J.F.Colombeau, ( jf.colombeau@wanadoo.fr).

Abstract. We present numerical techniques based on generalized functions adapted to nonlinear calculations. They concern main numerical engineering problems ruled by - or issued from - nonlinear equations of continuum mechanics. The aim of this text is to invite the readers in applying these techniques in their own work without significant prerequisites by presenting their use on a sample of elementary applications from engineering. Pure mathematicians can read it easily since the numerical techniques are fully recalled in a very simple way and these applied mathematics are indeed based on nonlinear pure mathematics, namely on the use of generalized functions adapted to nonlinear calculations. The "nonlinear generalized functions" permit to compute explicit solutions of simple problems, called Riemann problems, in cases in which this is not possible using classical generalized functions (such as the distributions). These explicit solutions are the elementary building blocks of efficient numerical schemes needed by engineers. This text was prepared for an engineering meeting. For mathematicians it is an illustration of general introductions to these "nonlinear generalized functions". For physicists it complements, concerning continuum mechanics, the use of the nonlinear generalized functions in general relativity.

**1-Prerequisites on generalized functions**. We shall use the Heaviside function H and the Dirac function $\delta$. They are defined by: $H(x) = 0$ if $x < 0$, $H(x) = 1$ if $x > 0$ (and $H(0)$ unspecified), $\delta(x)=0$ if $x \neq 0$, $\delta(0)$ "infinite" so that $\int \delta(x)dx = 1$. H is not differentiable at $x = 0$ in the classical sense because of the discontinuity there. However the Dirac function $\delta$ is considered as its derivative, in the sense that, $\forall \varphi$ infinitely differentiable function with compact support:

$$\int_{-\infty}^{+\infty} H(x)\varphi'(x)dx = \int_{0}^{+\infty} \varphi'(x)dx = -\varphi(0) = -\int_{-\infty}^{+\infty} \delta(x)\varphi(x)dx,$$

that is, $\delta$ plays the role of H' in the integration by parts formula. We shall use "nonlinear generalized functions" (ie. generalized functions adapted to nonlinear calculations) that the reader does not really need to know since the way to use them fits closely to physical and numerical intuition. In this viewpoint the book [C4] was written for engineers. The interested reader can look at the very elementary introductions [C1,C2,C3,G] for a wide audience and at the surveys [O1,S-V] on the use of nonlinear generalized functions in wave equations and general relativity respectively.

If $\Omega$ is an open set in $\mathbb{R}^N$ <u>the generalized functions on $\Omega$ are treated as if they were familiar $C^\infty$ functions that can be differentiated and multiplied freely, but there is a novelty.</u> Classically it is natural to state H²=H. In the present context this is false: <u>one has H²≠H</u>, although $\forall \varphi \in C_c^\infty(\Omega)$ (i.e. infinitely differentiable with compact support in $\Omega$) $\int(H^2(x)-H(x)).\varphi(x)dx$ is "infinitely small in absolute value", what we call "infinitesimal", a concept that fits with the familiar intuitive concept of "infinitesimal quantity" in physics. Let us explain this basic point in detail. In engineering the Heaviside function H represents a function whose values jump from 0 to 1 in a tiny interval around x=0; take this interval as $[0,\varepsilon]$: it is obvious that $\int(H^2(x)-H(x)).\varphi(x)dx$ tends to when $\varepsilon \to 0^+$ if $\varphi$ is a bounded function. But $\delta$ is unbounded: if H is represented by a straight line junction on $[0,\varepsilon]$, from the value 0 to the value 1, $\delta=H'$ has the value $= 1/\varepsilon$ if $0<x<\varepsilon$. Finally one has $\int(H^2(x)-H(x)).H'(x)dx =-1/6$, as obvious from elementary calculations.

    This shows that one is not allowed to state H²=H in a context in which the function H²-H should be multiplied by a function "taking infinite values" such as $\delta$. Therefore in a context of nonlinear generalized functions one is forced to distinguish functions that are "infinitesimal nonzero" such as H²-H, from the genuine zero function, because "infinitesimal nonzero quantities", when multiplied by "infinitely large quantities" can give nonzero results. For this we introduce a new notation: for 2 generalized functions $G_1$ and $G_2$ on an open set $\Omega$ we note $G_1 \approx G_2$ ("$G_1$ <u>associated to</u> $G_2$") if $\forall \varphi \in C_c^\infty(\Omega)$ (i.e. infinitely differentiable with compact support in $\Omega$) the number $\int_\Omega (G_1-G_2)(x) \varphi(x)dx$ is "infinitesimal" (i.e. it depends on a small parameter $\varepsilon$ and it tends to 0 as $\varepsilon \to 0^+$). The nonlinear generalized functions on $\Omega$ form a « differential algebra » denoted by $\mathbf{G}(\Omega)$ (i.e. one has differentiation and multiplication with their usual properties), see [C1,C2,C3]. $\mathbf{G}(\Omega)$ contains the classical functions and the "classical generalized functions" (called "distributions"), but the classical equality becomes either $=$ or $\approx$, according to the situation: for instance one has H²$\approx$H and H²$\neq$H. Not only it is not useful and obviously impossible to keep classical equalities such as H²=H but also physics displays instances in which several different Heaviside functions (associated to each other) are requested to model different physical variables: see the example of elastoplastic shock waves in figure 6 below and [C4 p106-107,C5].

    To summarize one computes on nonlinear generalized functions following the classical calculations on infinitely differentiable functions (using =), but a basic novelty lies in that <u>the classical equality of functions splits into = (the "true" or "strong" or "algebraic" equality) and the association $\approx$ (also called "weak" equality).</u> Concerning the association the basic points are that $G_1 \approx G_2$ <u>does not imply automatically</u> $G.G_1 \approx G.G_2$ if G is another generalized function in $\mathbf{G}(\Omega)$, but $G_1 \approx G_2$ implies $\frac{\partial}{\partial x_i} G_1 \approx \frac{\partial}{\partial x_i} G_2$. As an example H²$\approx$H but H²H' is not associated to HH': indeed H²$\approx$H implies by differentiation 2HH'$\approx$H' and $H^3 \approx H$ implies 3H²H'$\approx$H'. The association is some kind of weak equality not coherent with multiplication and strictly weaker than the (new) equality = in $\mathbf{G}(\Omega)$, but that often takes the place of the classical equality. Keeping this in mind the calculations on nonlinear generalized functions are very easy all the more so as they follow the physical intuition. The purpose of the sequel of the paper is to explain the above: choice between = and $\approx$, <u>perform calculations and use the result of these calculations to build Godunov numerical schemes on various significant problems of numerical engineering that cannot be treated without using nonlinear generalized functions.</u>

**2-Prerequisites on a few techniques of numerical engineering.**
**\*Standard 1D-notation.** We set as usual $\mathbf{x}_i = \mathbf{i.h}$, $\mathbf{h}>0$ is the space step, and $\mathbf{t}_n = \mathbf{n.}(\Delta \mathbf{t})$, $\Delta \mathbf{t}>0$ is the time step; i and n are integer numbers, n$\geq$0 and $-\infty<i<+\infty$, limited by the size of the calculations; $\mathbf{r} =(\Delta \mathbf{t})/\mathbf{h}$ <u>is the CFL number</u> (Courant-Friedrichs-Levy). Therefore space-time

is divided into cells $\{i.h \leq x \leq (i+1).h\} \times \{n.(\Delta t) \leq t \leq (n+1).(\Delta t)\} = [x_i, x_{i+1}] \times [t_n, t_{n+1}]$. The solution U of the system of equations under consideration is supposed to be known at time $t_n$, constant on each space interval $\{i.h < x < (i+1).h\}$, thus providing the set of numbers $\{U^n_{i+1/2}, i \in \mathbf{Z}\}$ (where $\mathbf{Z}$ denotes of course only the finite set of integers under consideration, and each $U^n_{i+1/2}$ is a set of p real numbers in case of a system of p scalar equations: $U(x,t) = U^n_{i+1/2}$ if $i.h < x < (i+1).h$ and $t = t_n$). One seeks the set of numbers $\{U^{n+1}_{i+1/2}, i \in \mathbf{Z}\}$ proceeding by induction on n until the desired final time.

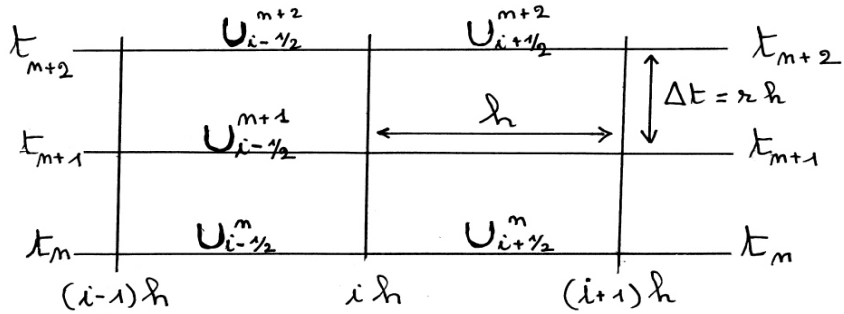

Figure 1. The grid and the notations.

 **The Godunov 1D-method.** We shall use the "Godunov method" because of its high numerical quality: numerical schemes from "trivial" discretizations need "artificial viscosity" which makes them degenerate too quickly. The Godunov method consists first in solving the "Riemann problems" at all points $x_i = i.h$ (at time $t_n$) from the knowledge of the left hand side values $U^n_{i-1/2}$ and the right hand side values $U^n_{i+1/2}$. The "Riemann problem" is the particular initial value problem when the data are constant on both sides of a discontinuity. In practice for the equations under consideration we solve the Riemann problem by constant states separated by discontinuities moving at constant speed (examples in [C4]). We describe the method on a standard example which makes sense within the "distributions" (i.e. classical generalized functions), but this is unimportant for the sequel. Consider the scalar equation (for which the Riemann problem can be solved classically using distributions: the symbol = in (1) is intended in the sense of distributions because this is more familiar to the expert reader; lower indices denote partial derivatives):

(1) $\quad u_t + 1/2.(u^2)_x = 0.$

Inserting the formula

(2) $\quad u(x, t) = u_l + (u_r - u_l).H(x - ct)$

into (1), and stating here, exceptionally at this point, H²=H because this is = in the sense of distributions (which is possible here because no further multiplication occurs) one finds at once that (2) is solution of (1) if and only if

(3) $\quad c = (u_r + u_l)/2.$

Remark. The **G**-context - to be used for equations for which discontinuous solutions do not make sense within the distributions - is completely coherent with distribution theory and one obtains at once the same jump conditions (when they can be obtained within the distributions as this is the case of (1)): see Remark 1 in section 9 below. In the **G**-context (i.e. in $\mathbf{G}(\Omega)$ for the open set $\Omega$ under consideration) (1) is stated as $u_t + 1/2.(u^2)_x \approx 0$; one has $H^2 \approx H$, therefore (H²)' ≈ H', and one obtains (3) again from same calculations with ≈ in place of = in the sense of distributions.

Therefore the solution of the Riemann problem is made of one discontinuity travelling at velocity c and thus it depends on the sign of c. Below we draw the 4 aspects of the solutions of the 2 Riemann problems issued from the points $(x_i, t_n)$ and $(x_{i+1}, t_n)$ according to the signs of $c_i^n = (u_{i-1/2}^n + u_{i+1/2}^n)/2$ and $c_{i+1}^n = (u_{i+1/2}^n + u_{i+3/2}^n)/2$:

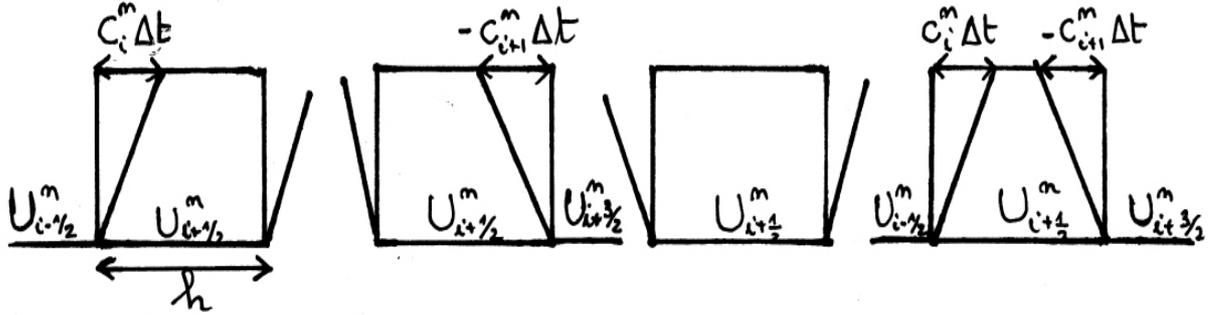

Figure 2. From left to right: $c_i^n \geq 0$ and $c_{i+1}^n \geq 0$, $c_i^n \leq 0$ and $c_{i+1}^n \leq 0$, $c_i^n \leq 0$ and $c_{i+1}^n \geq 0$, $c_i^n \geq 0$ and $c_{i+1}^n \leq 0$.

The CFL number r is supposed small enough so that the discontinuities issued from the points $x_i$ and $x_{i+1}$ at time $t_n$ do not meet within the cell $[x_i, x_{i+1}] \times [t_n, t_{n+1}]$. Then the second step in the Godunov method consists in taking as $u_{i+1/2}^{n+1}$ the average of the solution thus obtained on $[x_i, x_{i+1}]$ at time $t_{n+1}$: this is called the projection step. For the above problem in each cell $[x_i, x_{i+1}] \times [t_n, t_{n+1}]$ one has 4 possibilities according to the signs of the velocities stemming from $x_i$ and $x_{i+1}$. The CFL condition $r.\max(|c_i^n|) \leq 1/2$ ensures the discontinuities cannot meet inside the cells and one obtains the following formulas in the 4 successive cases in fig.2 from left to right:

$$u_{i+1/2}^{n+1} = r.c_i^n . u_{i-1/2}^n + (1 - r.c_i^n).u_{i+1/2}^n;$$
$$u_{i+1/2}^{n+1} = (1 + r.c_{i+1}^n).u_{i+1/2}^n - r.c_{i+1}^n.u_{i+3/2}^n;$$
$$u_{i+1/2}^{n+1} = u_{i+1/2}^n;$$
$$u_{i+1/2}^{n+1} = r.c_i^n . u_{i-1/2}^n + (1 - r.c_i^n + r.c_{i+1}^n).u_{i+1/2}^n - r.c_{i+1}^n.u_{i+3/2}^n.$$

For a system with more than one scalar equation the solution of a Riemann problem is made of several discontinuities that stem from the point of discontinuity of the initial condition. It becomes already very complicated for 4 scalar equations or more, see [C4,LR1].

**\*Splitting of equations.** The symbol "=" in equations (4) to (9) is intended in a "non clarified" or "formal" sense when nonclassical products occur (in the **G**-setting the point will be to replace it by = in **G** or by ≈, as explained in [C1,C2] and in section 3 below). Consider an equation

(4)     $U_t + f(U, U_x) + g(U, U_x)$ "=" 0.

One starts with the set of values $\{U_{i+1/2}^n, i \in \mathbf{Z}\}$. The splitting of equations consists in considering, from time $t_n$ to $t_{n+1}$, first the equation

(5)     $U_t + f(U, U_x)$ "=" 0,

in order to obtain, by the Godunov method (or another method), values denoted by $U_{i+1/2}^{n+1/2}$ (the upper index n+1/2 is purely conventional to note an intermediate value: one considers the final time is really $t_{n+1}$). Then, starting from the set of values $\{U_{i+1/2}^{n+1/2}, i \in \mathbf{Z}\}$, considered as values at time $t_n$, one considers (again from time $t_n$ to $t_{n+1}$), the equation

(6)    $U_t + g(U,U_x)$ "="0

in order to obtain the desired set of values $\{U_{i+1/2}^{n+1}, i \in \mathbf{Z}\}$. Experience has proved the results so obtained are often quite good and this splitting method has an enormous advantage: the Godunov method is based on usually difficult algebraic calculations: those for the solutions of the Riemann problems. These calculations might be impossible for (4) (or extremely difficult), and easy for both (5) and (6), with possible numerical advantages.

**Dimensional splitting.** In 2D (denoting the variables x and y) consider the equation

(7)    $U_t + f(U,U_x) + g(U,U_y)$ "="0.

The dimensional splitting consists in studying successively from time $t_n$ to $t_{n+1}$ the equations (8),(9) below which separate the x and y directions: first

(8)    $U_t + f(U,U_x)$ "="0,

whose result is taken as initial data for

(9)    $U_t + g(U,U_y)$ "="0.

The advantage is to permit the use of the Godunov method above (which is a 1D-method because of the too great complexity of the solution of 2D-Riemann problems) for each system (8) and (9).

### 3-A first use of generalized functions: statement and transformation of equations.

**Physical interpretation.** In [C2] we showed that the statement $u_t + uu_x = 0$ with = in **G** is erroneous when one considers shock waves and that the correct statement is: $u_t + uu_x \approx 0$. Consider the system below or the one in top of section 4 which arise directly from physics: one distinguishes between basic laws (conservation of mass, momentum, total energy) which are same for all materials, and constitutive equations (also called state laws) that make the distinction between different materials and are issued from specific experiments (they appear in form of empirical formulas, and more generally of arrays of numbers that are experimental data). Consider a shock wave: it is a jump in the values of the physical variables. It is known to take place on a small space interval (in the direction of propagation) of the order of length of, may be, about 100 times the average distance between atoms or molecules. In this region one may consider by thought small bags (small cubes dx.dy.dz) during a time dt in which we can state the basic conservation laws. Since the width of the shock waves appears to be the standard infinitesimal length under consideration this suggests to state the basic laws with = in **G**. On the other hand the state laws have never been observed in a state of fast deformation such as the inside of a shock wave, which suggests to state them with the association $\approx$. In 2D (easy extension to 3D) this gives the following statement of the system of fluid dynamics in **G** (with = and $\approx$):

1. $\rho_t + (\rho u)_x + (\rho v)_y = 0$        mass conservation
2. $(\rho u)_t + (\rho u^2)_x + (\rho u v)_y + p_x = 0$        momentum conservation in x-direction
3. $(\rho v)_t + (\rho u v)_x + (\rho v^2)_y + p_y = 0$        momentum conservation in y-direction
4. $(\rho e)_t + (\rho e u)_x + (\rho e v)_y + (p\, u)_x + (p\, v)_y = 0$        energy conservation

5. $\quad p \approx \Phi(\rho, e - (u^2 + v^2)/2)$   state law (also called constitutive equation);

$\Phi$ is a function issued from experiments.

**\*Mathematical confirmation:** Since we are interested in shock waves solutions the first question is: does this formulation allow such solutions (remind the example $u_t + uu_x = 0$)? This system has been studied (in 1D which is OK since we are interested in such waves and further since the equations do not depend on a particular direction) in [C4 §4.5] and the answer is yes.

**\*Application of generalized functions to simplification of equations**: On the first 4 equations one can compute as usual (i.e, as in case of $C^\infty$ functions) because they are stated with = in **G**. Inserting 1. into 2.,3.,4. and setting $\lambda = 1/\rho$ (assuming $\rho$ is invertible in **G,** i.e. no void region), give respectively ($T = e - (u^2 + v^2)/2$ is the temperature =internal energy):

1. $\quad \lambda_t + u\lambda_x + v\lambda_y = \lambda.(u_x + v_y)$
2. $\quad u_t + uu_x + vu_y = -\lambda p_x$
3. $\quad v_t + uv_x + vv_y = -\lambda p_y$
4. $\quad T_t + uT_x + vT_y = -\lambda p.(u_x + v_y)$
5. $\quad p \approx \Phi(1/\lambda, T)$.

These equations are valid even in case of shock waves which is not trivial in view of the above remark on eq. $u_t + uu_x = 0$ and the occurrence of this disappointing event as a mistake in many formal calculations done by engineers. The dimensional splitting (here the system in x-direction) gives:

1. $\quad \lambda_t + u\lambda_x = \lambda u_x$
2. $\quad u_t + uu_x = -\lambda p_x$
3. $\quad v$ =constant (from $v_t + uv_x = 0$)
4. $\quad T_t + uT_x = -\lambda p u_x$
5. $\quad p \approx \Phi(1/\lambda, T)$.

The jump conditions are the (usual) Rankine-Hugoniot formulas since this last system is equivalent to the classical 1D fluid dynamics system. Therefore the solution of the Riemann problem is known but quite complicated. This system can be also treated (with some advantages) by splitting of equations as done in section 11 below before figure 5, using the nonlinear generalized functions to obtain  <u>far simpler solutions of the Riemann problem for the systems issued from the splitting of equations.</u>

**4-Collisions of solids with strong deformations (cartesian coordinates).** First notice that for strong collisions the Lagrangian representation cannot be used in numerical calculations because of large deformations of the grid. All equations below are given in Eulerian representation. In 2D engineers use the following system of 8 equations for 8 unknown functions $\rho, u, v, p, e, s_{11}, s_{12}, s_{22}$ ($\rho$ = mass per unit volume, $(u,v)$ = velocity vector, $p$ = pressure, $e$ = total energy per unit volume, ($s_{ij}$) = stress tensor):

1. $\quad \rho_t + (\rho u)_x + (\rho v)_y = 0$
2. $\quad (\rho u)_t + (\rho u^2)_x + (\rho uv)_y + (p - s_{11})_x - (s_{12})_y = 0$
3. $\quad (\rho v)_t + (\rho uv)_x + (\rho v^2)_y + (p - s_{22})_y - (s_{12})_x = 0$
4. $\quad (\rho e)_t + (\rho eu)_x + (\rho ev)_y + [(p - s_{11})u]_x + [(p - s_{22})v]_y - (s_{12}v)_x - (s_{12}u)_y = 0$

5. $(s_{11})_t + u.(s_{11})_x + v.(s_{11})_y \approx 4/3.\mu(\text{--}).u_x - 2/3.\mu(\text{--}).v_y + (u_y - v_x).s_{12}$

6. $(s_{22})_t + u.(s_{22})_x + v.(s_{22})_y \approx -2/3.\mu(\text{--}).u_x + 4/3.\mu(\text{--}).v_y - (u_y - v_x).s_{12}$

7. $(s_{12})_t + u.(s_{12})_x + v.(s_{12})_y \approx \mu(\text{--}).v_x + \mu(\text{--}).u_y - 1/2.(u_y - v_x).(s_{11} - s_{22})$

8. $p \approx \Phi(\rho, e - (u^2 + v^2)/2)$,

where $\mu(\text{--})$ is a function of the variables $s_{ij}$ ($\mu$ is the "shear modulus": constant in the elastic stage, null in the plastic stage) and $\Phi$ is a function. Both functions $\mu$ and $\Phi$ are obtained from experiments on the solid under consideration. Equations 1-2-3-4 are the basic conservation laws (stated with = in **G**) while 5-6-7-8 are state laws (stated with ≈), as explained above and in [C1,C2]. All terms in eqs. 5,6,7 except the t-derivatives involve products of an Heaviside function and a Dirac delta function which do not make sense in distribution theory and so require the **G**-context. This 2D system is treated by dimensional splitting, giving rise to 2 simpler systems of 1D equations that still involve the same products that require the **G**-context. A use of the **G**-context to solve the Riemann problem for these equations is based on the natural interpretation of the equations of physics with = in **G** and ≈ as described in section 3 and [C1,C2]. Since there are many equations stated with the association one needs some additional information to resolve the remaining ambiguity (problem in section 11 below). A 1D version of the above system ($s = s_{11}$, $s_{12} = 0 = s_{22}$) is:

1. $\rho_t + (\rho u)_x = 0$
2. $(\rho u)_t + (\rho u^2)_x + (p - s)_x = 0$
3. $(\rho e)_t + (\rho e u)_x + [(p - s)u]_x = 0$
4. $s_t + u.(s)_x \approx 4/3.\mu(\text{--}).u_x$
5. $p \approx \Phi(\rho, e - u^2/2)$.

This system is a simplification of those obtained by dimensional splitting from the above 2D system. It is studied in section 11 below using a splitting of equations; as an application of the nonlinear generalized functions:

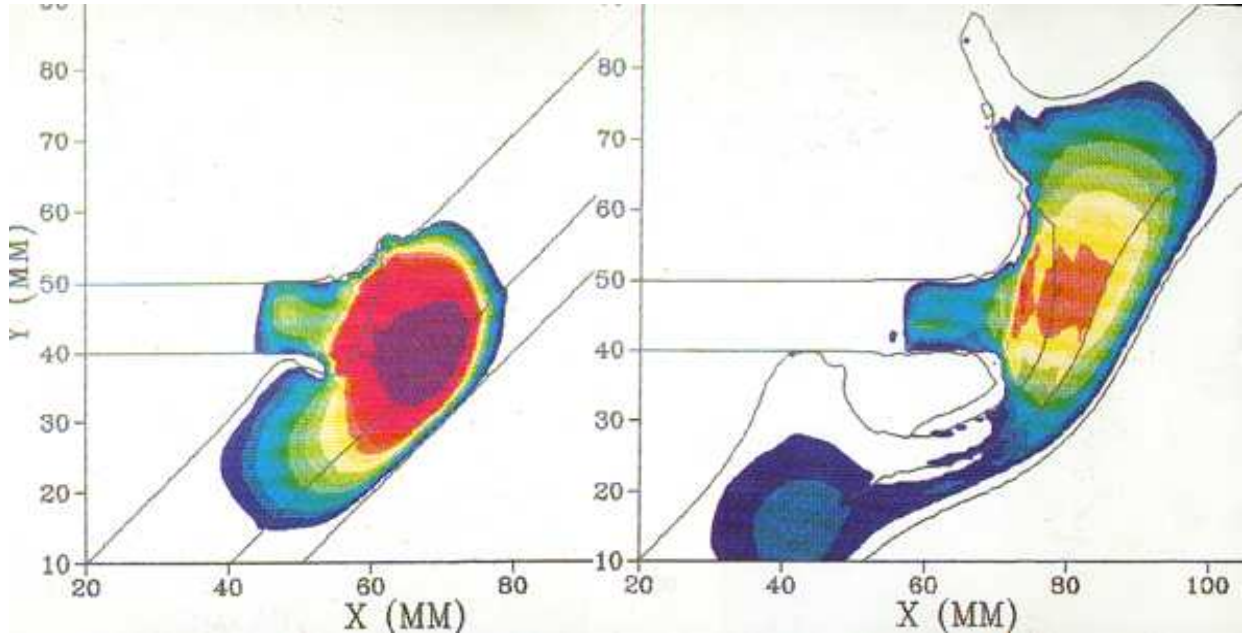

Figure 3. A numerical solution of the above 2D system 5 and 15 microseconds after the collision (explanations are given in [C2]), obtained using the dimensional splitting, then

splittings of the 1D systems and the Godunov method. <u>The Godunov method is based on the use of nonlinear generalized functions developed in sections 9 and 11 below to obtain the needed solutions of the Riemann problems.</u>

**5-Collisions of solids with strong deformations (cylindrical coordinates).** The 3D-system of continuum mechanics is often used in cylindrical coordinates to model axisymmetric collisions. Coordinates: r,z (an axisymmetric phenomenon is independent on the $\theta$ coordinate). The lower indices are partial derivatives; the upper indices denote different physical variables. The axisymmetric system is given below ((u,v)=components of the velocity perpendicular and parallel to the z-axis respectively; the terms divided by r should have a limit at $r \to 0$). Equations 1 to 4 have been simplified using = in **G** that permits to calculate as in case of usual functions.

1. $\rho_t + \rho.u_r + \rho.v_z = -\rho.u/r$
2. $\rho.u_t + (p - s^{rr})_r - (s^{rz})_z = (2s^{rr} + s^{zz})/r$
3. $\rho.v_t + (p - s^{zz})_z - (s^{rz})_r = s^{rz}/r$
4. $\rho.e_t + [u.(p - s^{rr}) - v.s^{rz}]_r + [v.(p - s^{zz}) - u.s^{rz}]_z = [u.(-p + s^{rr}) + v.s^{rz}]/r$
5. $(s^{rr})_t - 4/3.\mu(-\!-).u_r + 2/3.\mu(-\!-).u_z - (u_z - v_r).s^{rz} \approx -2/3.\mu(-\!-).(u/r)$
6. $(s^{zz})_t + 2/3.\mu(-\!-).u_r - 4/3.\mu(-\!-).u_z + (u_z - v_r).s^{rz} \approx -2/3.\mu(-\!-).(u/r)$
7. $(s^{rz})_t - \mu(-\!-).(u_z + u_r) + 1/2.(u_z - v_r).(s^{rr} - s^{zz}) \approx 0$
8. $p \approx \Phi(\rho, e - (u^2 + v^2)/2)$.

This system is as usual treated first by dimensional splitting, then by splitting of equations and by the use of nonlinear generalized functions exposed in sections 9 and 11 below [LR1]:

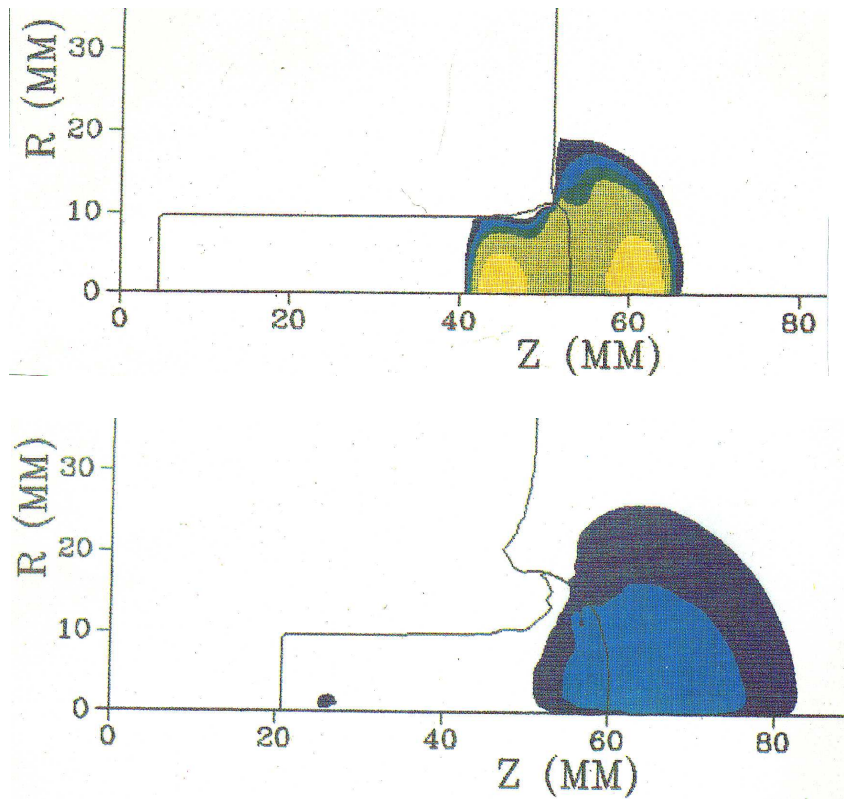

Figure 4. A numerical solution: a collision steel-steel at 2 and 10 microseconds after the collision; speed of the projectile: 2000m/s; similar explanations as in the case of fig.3, see [C2].

**6-Multifluid flows.** We consider a mixture of n fluids made of bubbles of one fluid into the other such as nonmiscible fluids that have been strongly shaken. If we admit that there is only one pressure $p$ in the mixture (i.e. the pressure at a point of the individual fluids are the same) we obtain the following 1D-system: $\alpha_i$(x,t) denotes the volumic proportion of fluid i in the mixture, $1 \leq i \leq n$; $\rho_i$ = mass of fluid i per unit volume of fluid i, $u_i$ = velocity of fluid i, $e_i$ = total energy of fluid i per unit volume of fluid i, $\Phi_i$ = state law of fluid i.

1. $(\alpha_i \rho_i)_t + (\alpha_i \rho_i u_i)_x = 0$
2. $(\alpha_i \rho_i u_i)_t + (\alpha_i \rho_i (u_i)^2)_x + \alpha_i . p_x = 0$
3. $(\alpha_i \rho_i e_i)_t + [(\rho_i e_i + p).\alpha_i .u_i]_x + p.(\alpha_i)_t = 0$
4. $p \approx \Phi_i(\rho_i, e_i - (u_i)^2/2)$
5. $\sum_{i=1}^{n} \alpha_i = 1.$

This gives 4n+1 equations for 4n+1 unknowns. The equations in 2D and 3D are given in [S-W].The nonclassical products are in the last terms of eqs. 2 and 3. Numerical results are in [B-C,C4].

**7-Products arising from a splitting of the system of fluid dynamics.** In the above models the products requesting the **G**-context arose directly from the models given by engineers and physicists. The classical system of fluid dynamics does not show such products: here are the equations in 1D (same notation as above):

1. $\rho_t + (\rho u)_x = 0$
2. $(\rho u)_t + (\rho u^2)_x + p_x = 0$
3. $(\rho e)_t + (\rho e u)_x + (pu)_x = 0$
4. $p \approx \Phi(\rho, e - u^2/2).$

One says it is in conservation form and therefore the solution of the Riemann problem has been obtained since long time (from the classical Rankine-Hugoniot jump conditions). For some applications [LR1] it appears useful to treat it by the following splitting of equations that needs the **G**-context [LR1,B-B-D-LR]; <u>first step</u>:

1. $\rho_t + (\rho u)_x \approx 0$
2. $(\rho u)_t + (\rho u^2)_x \approx 0$
3. $(\rho e)_t + (\rho e u)_x \approx 0;$

<u>second step</u>:

1. $\rho$ = constant  (from $\rho_t = 0$)
2. $(\rho u)_t + p_x = 0$
3. $(\rho e)_t + (pu)_x = 0$
4. $p \approx \Phi(\rho, e - u^2/2).$

The first system shows a solution of the Riemann problem in which there appears a Dirac delta function in the variable $\rho$ while u is discontinuous, see [B-B-D-LR,B-C-M,Hu ], thus the product of $\rho$ and u does not make sense classically. For other "delta shocks" see [C4,J1,J3,J4,N,N-O].

**8-A simple model for hurricanes**. A model issued from the system in section 3 (fluid dynamics) has been used in [LR-LR-M] as a simple model for hurricanes. The basic role is played by the horizontal components of the velocity vector (u,v). It is stated as:

1. $\qquad u_t + uu_x + vu_y = (k-\mu)(u-u_*) + \omega(v-v_*)$
2. $\qquad v_t + uv_x + vv_y = (k-\mu)(v-v_*) - \omega(u-u_*)$,

$\omega$ =Coriolis coefficient; $\mu \geq 0$ and $k \geq 0$ are coefficients modelling friction and the action of the vertical velocity respectively; $(u_*,v_*)$ is a field of velocity corresponding to the trade winds; these coefficients can be actualized in time from observations. There are discontinuities of u and v on the boundary of the hurricane and on the wall of the eye; then the terms $vu_y$ and $uv_x$ show nonclassical products which are contact discontinuities. This case is not treated by dimensional splitting, but by integration along characteristic curves, which is possible in this simple case, see[LR-LR-M]: one observes the creation and evolution of a structure in form of ring with a stable hole in the center (eye of the hurricane) and one can observe the trajectory of the hurricane.

<u>Remark: a shallow water model</u> [LR2]. h=h(x,t), a=a(x) smooth or not, g is the gravity constant. The model is:

$$h_t + (hu)_x = 0$$
$$(hu)_t + (hu^2)_x + gh(h+a)_x = 0.$$

A numerical study is in [LR2] and references there.

**9-Construction of numerical schemes of the Godunov type: an example.** After this list of a few samples of systems of PDEs (simplified due to the nature of this expository paper) used by engineers, the main point is to obtain jump formulas (replacing for these systems the classical jump formulas of systems that do not show nonclassical products). One chooses a proper statement of the equations in the **G**-context, as described in section 3 and [C1,C2], by properly using = (in **G**) and $\approx$ on physical ground.

*****Statement of a system of 3 equations.** We consider the system in [C1,C2] (it is explained there how = and $\approx$ have been chosen on physical ground, see also section 3).

$$\rho_t + (\rho u)_x = 0$$
$$(\rho u)_t + (\rho u^2)_x = \sigma_x$$
$$\sigma_t + u\sigma_x \approx k^2 u_x, \ k>0.$$

We seek formulas for a shock wave:

$\qquad$ (sw) $\qquad w(x,t) = w_l + \Delta w H_w(x-ct), \quad \Delta w = w_r - w_l, \quad w = \rho, u, \sigma, \rho u, \rho u^2$.

*****Classical jump formulas for the first 2 equations.** Even if stated with the association (i.e. in a weaker form: this suffices here because they are in conservation form: shock waves make sense within the distributions) they give:

$\qquad$ -c $\Delta\rho(H_\rho)' + \Delta(\rho u)(H_{\rho u})' \approx 0$, -c $\Delta(\rho u)(H_{\rho u})' + \Delta(\rho u^2)(H_{\rho u^2})' \approx \Delta\sigma(H_\sigma)'$.

Since all Heaviside functions are associated (therefore their derivatives are associated to any Dirac delta function) one has the following formulas involving the coefficients:

$$c = \frac{\Delta(\rho u)}{\Delta\rho}, \ c = \frac{-\Delta\sigma + \Delta(\rho u^2)}{\Delta(\rho u)};$$

Setting $v=1/\rho$, these formulas will be used in the form (easy classical calculations):

$$(jf1\&2) \qquad c - u_l = -v_l \frac{\Delta u}{\Delta v}, \quad (\Delta u)^2 = \Delta\sigma.\Delta v.$$

<u>Remark 1</u>: These are nothing else than the classical jump formulas obtained from the fact the 2 equations under consideration are "in conservation form". As a consequence, up to now, we did not use the fact these 2 equations are indeed stated with = in **G**, but only with $\approx$.

<u>Remark 2</u>: This fact that the first 2 equations are in conservation form has brought a simplification in calculations. <u>However this simplification is unessential when one uses the fact they are stated with = in **G**: if these equations had been in nonconservation form, their statement with = in **G** would have given as well nonambiguous jump conditions, but the calculations would have been more complicated: see section 12.</u>

**\*Exploitation of the statement of the first 2 equations with = in G to compare the Heaviside functions in the shock**. Using $v=1/\rho$ (v was denoted $\lambda$ in section 3; here we find more convenient to follow the notation of [C4] where more details can be found) the statement of these equations with = in **G** permits to transform them into (immediate calculations):

$$(teq) \qquad v_t + uv_x - vu_x = 0, \quad u_t + uu_x - v\sigma_x = 0.$$

Inserting (sw) into the first equation in (teq) one obtains:

$$(c - u_l - \Delta u H_u).(H_v)' = -(\frac{v_l}{\Delta v} + H_v).\Delta u.(H_u)'.$$

Using the first equality in (jf1&2):

$$(eq1) \qquad (\frac{v_l}{\Delta v} + H_u).(H_v)' = (\frac{v_l}{\Delta v} + H_v).(H_u)'.$$

For the second equation in (teq):

$$(c - u_l - \Delta u H_u).\Delta u.(H_u)' = -(v_l + \Delta v H_v).\Delta\sigma(H_\sigma)'.$$

Using (jf1&2):

$$(-v_l \frac{\Delta u}{\Delta v} - \Delta u H_u).\Delta u.(H_u)' = -(v_l + \Delta v H_v)\frac{(\Delta u)^2}{\Delta v}(H_\sigma)'$$

$$(eq2) \qquad (\frac{v_l}{\Delta v} + H_u).(H_u)' = (\frac{v_l}{\Delta v} + H_v).(H_\sigma)'.$$

To shorten the notation set $\alpha = \frac{v_l}{\Delta v}$. Then (eq1 & 2) give

$$(\alpha + H_u)(H_v)' = (\alpha + H_v)(H_u)' , \quad (\alpha + H_u)(H_u)' = (\alpha + H_v)(H_\sigma)';$$

The first equation can be stated as: $(\alpha + H_u).(H_v)' - H_u'.H_v - \alpha.H_u' = 0$. This is an equation of the familiar kind a(x)y' + b(x)y +c(x)=0 with y= $H_v$. In the **G**-context it can be treated as usual since stated with = in **G**. Finally one finds $H_v = H_u$. Using the second equation one finally obtains that $H_v = H_u = H_\sigma$. Note that such an unexpected simple result is due to this particular system and this particular choice of physical variables: for instance $H_\rho$ is different from them.

**\*The 3 jump formulas.** We have already obtained the 2 jump formulas (jf1&2) - which here are in fact the classical ones. Now we can obtain the (nonclassical) jump formula for the third eq. $\sigma_t + u\sigma_x \approx k^2 u_x$. Noting $H$ the Heaviside function $H_v = H_u = H_\sigma$ and putting it into the third equation:

$$-c\Delta\sigma H' + u_l \Delta\sigma H' + \Delta u.\Delta\sigma.H.H' \approx k^2 \Delta u H'.$$

Since $H^2 \approx H$ one has $2HH' \approx H'$ and therefore the third jump formula is:

(jf3) $\quad -c\Delta\sigma + u_l\Delta\sigma + \Delta u.\Delta\sigma/2 = k^2\Delta u$ .

Combining (jf1&2) with (jf3) gives:

(jf3') $\quad v_l \dfrac{\Delta\sigma}{\Delta u} = -\dfrac{\Delta u}{2} + k^2 \dfrac{\Delta u}{\Delta\sigma}$ .

The jump formulas (jf1&2, jf3') permit to find the solution of the Riemann problems at the junctions of meshes (this is a difficult algebraic calculation already for a system of 3 equations, see [C4,LR1]), and then to build a Godunov type scheme as explained in section 2, see [C4, LR1] for a number of examples of these calculations.

**10-Mathematical solutions and convergence.** In a few simple cases (systems of 2 equations) one can obtain suitable estimates on the Godunov type numerical schemes so constructed and from them, by compactness, prove existence of a solution in the association sense and convergence of a subsequence of approximate solutions. See [A-C-LR, C-C-LR, Bia, B-C-M-R, M-P,J2].

**11-Construction of numerical schemes of Godunov type: another example.** We consider the 1D-system of elastoplasticity in section 4:

1. $\quad \rho_t + (\rho u)_x = 0$
2. $\quad (\rho u)_t + (\rho u^2)_x + (p - s)_x = 0$
3. $\quad (\rho e)_t + (\rho e u)_x + [(p-s)u]_x = 0$
4. $\quad s_t + u.(s)_x \approx k^2(s)u_x$
5. $\quad p \approx \Phi(\rho, e - u^2/2)$.

with $k^2(s) = k^2$ if $|s| < s_0$ and $k^2(s) = 0$ if not, $k > 0$ and $s_0 > 0$. If $|s| < s_0$ the solid is elastic and if not it is plastic i.e. it behaves like a liquid (definitive changes in shape, breakings). The variable s represents the ability of the electrons to maintain the crystalline structure as long as $|s| < s_0$. For simplification we want to eliminate the variable $e$ so as to have only 4 equations and 4 unknown functions. We admit the state law 5. can be stated as $5'$: $e \approx \psi(\rho, p) + \dfrac{u^2}{2}$. We also assume we are only concerned with a solution which is regular except at individual points (shock waves or contact discontinuities). Outside these points the material is practically at rest and so the state law $5'$ can be stated with = in **G**, thus, outside these points, it can be put into eq. 3. Setting $v = \dfrac{1}{\rho}$, calculations in [C4 p75] permit the elimination of the variable $e$ outside the points of discontinuity. Taking into account the formula is obtained with = in **G** outside the (isolated) points of discontinuity, and therefore stating it globally with association, the above system is replaced by:

1. $\quad v_t + uv_x - vu_x = 0$
2. $\quad u_t + uu_x + v.(p-s)_x = 0$
3. $\quad s_t + us_x - k^2(s)u_x \approx 0$

4. a complicated equation see [C4 p76] which reduces to $p_t + up_x + \gamma p u_x \approx 0$ when the state law is $p \approx (\gamma-1)\rho.(e - \frac{u^2}{2})$, $\gamma > 1$. From the first 2 equations we know (section 9 above) that the jumps of the 3 variables $v, u, p-s$ are modelled by the same Heaviside function. There subsists an ambiguity in the jump conditions of eqs.3. and 4. (in $us_x, up_x, pu_x$). When $|s| \geq s_0$ the solid is "plastic" (i.e. "fluid") and eq.3. reduces to s=constant, therefore $H_v = H_u = H_p$ (the system reduces to the system of fluid dynamics). When $|s| < s_0$ the solid is "elastic" (i.e. maintained by the electrons) and one has 2 equations stated with $\approx$ (eq.3. and eq.4.). Then in the elastic case the calculations in section 9 give that the shocks of $v, u, p-s$ are all 3 modelled by the same Heaviside function. But the individual terms $us_x$ and $up_x$ (equivalently $pu_x$) request some link between the Heaviside functions of $u$ and $s$, and of $u$ and $p$ respectively. The needed information should stem from physics.

Problem: compare the 2 kinds of state laws: 5.6.7. in section 4 (a form of Hooke's law loosely stating that the deformation of an elastic solid is proportional to the force acting on it) and 8. in section 4 (relation between pressure, density and temperature), so as to find a proper information on the Heaviside functions modelling $p$ and $s$.

In absence of this information one has rather arbitrarily chosen to take as equal the Heaviside functions modelling $p$ and $s$ in the elastic part of a shock wave as a first approximation since both vary at the same time (although there is no indication they really are equal: anyway they are not very different). The results so obtained (in particular figures 3,4,5,6 - of course with the proper state laws of the solids under consideration) have been a posteriori considered as good (but the experimental verifications were not very precise, which leaves the door open for improvements).

The system in this section is treated in [LR1] (for the industrial application to the design of armor) using same Heaviside functions for $v, u, s, p$ and the splitting of equations:

first step:
1. $\qquad v_t + uv_x \approx 0$
2. $\qquad u_t + uu_x \approx 0$
3. $\qquad s_t + us_x \approx 0$
4. $\qquad p_t + up_x \approx 0$

and then second step:
1. $\qquad v_t - vu_x \approx 0$
2. $\qquad u_t + v.(p-s)_x \approx 0$
3. $\qquad s_t - k^2(s)u_x \approx 0$
4. $\qquad p_t + \gamma p u_x \approx 0$ .

Remark. A more precise treatment than in [LR1] is given in [M-P] taking into account a more precise solution of the Riemann problem in the first step (at the interfaces $x_i = ih$ such that $u_{i-1/2}^n \times u_{i+1/2}^n < 0$; in the industrial applications considered in [LR1] they do not play a significant role because the indices $i$ under concern are rare) in a particular case convenient for university studies.

One observes numerically known phenomena of the greatest interest to engineers such as "elastic precursors" and "elastoplastic shock waves". Figures 5 and 6 represent two numerical solutions of the Riemann problem modelling a 1D-collision (of a projectile from the left on a target at rest: $u$ is the velocity, $\sigma = s - p$, $\rho = 1/v$ is approximately constant and not represented) for the system on top of this section with two different sets of initial conditions that give qualitatively very different results[LR1]:

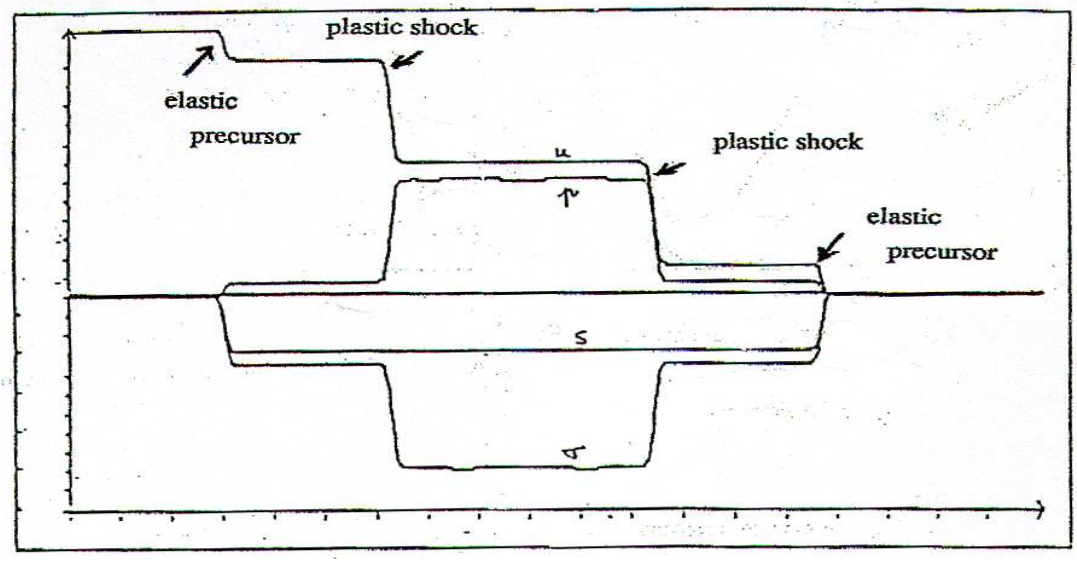

Figure 5: **Elastic precursors** followed by plastic shock waves (one pair moving to the left and another pair moving to the right). See [Bia p179-188] and see in [Bia p189,C5] an explanation (from elastic precursors) given by engineers concerning possible survival of some passengers in plane accidents when they take place in the domain of elastic precursors. The elastic precursors are not destructive but they put the material in an extremely fragile state in which it cannot forbear any further constraint (thus producing irremediable modifications or breakings). By modifying parameters, such as an increase in the speed of the collision, the speed of the plastic shock increases more than the one of the elastic precursor. When it reaches the speed of the elastic precursor they are glued together into an elastoplastic shock wave (fig.6) (that does not permit survival of any passenger as in the above case).

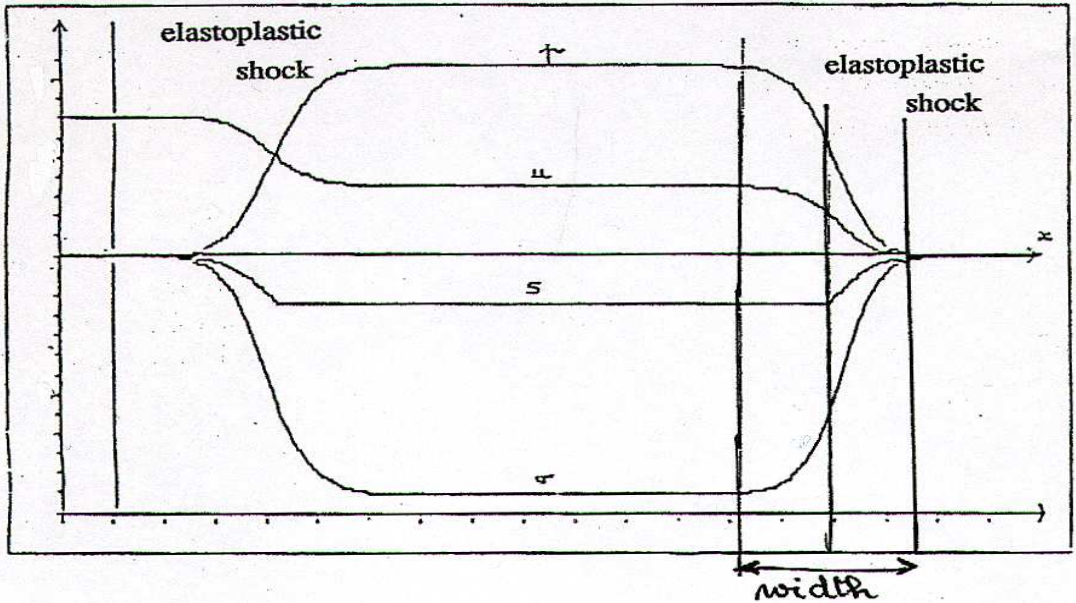

Figure 6: Two **elastoplastic shock waves** (one moving to the left and one moving to the right): the variable s varies only in a part of the shock while the variables p,u,$\sigma$ vary throughout the shock : they are modelled by very different Heaviside functions. This shows that **an infinity of different Heaviside functions is needed to model physics:** indeed the part of the shock waves in which the variable s remains constant (= plastic part: the electrons cannot maintain the crystalline structure) can be very small or can nearly cover the whole width according to the shock wave under consideration. See [Bia p179-188,C4,C5].

**12-An example of calculation of jump conditions** (remark 2 in section 9): Starting from the eq. $\sigma_t + u\sigma_x = u_x$ (stated here with = in **G** as a very simple mathematical example to show calculations) which is not in conservation form we show how we compute jump conditions. This calculation is based on the statement of the equation with = in **G**. Setting $u(x,t) = u_l + \Delta u H_u(x-ct)$, $\sigma(x,t) = \sigma_l + \Delta\sigma H_\sigma(x-ct)$ and putting these formulas in the equation one obtains:

$$(H_\sigma)' = \frac{\Delta u}{\Delta\sigma} \times \frac{(H_u)'}{-c + u_l + \Delta u H_u}.$$

Integrating from $-\infty$ to $+\infty$ and after the change of variable $\lambda = H_u$ it gives

$$1 = \frac{\Delta u}{\Delta\sigma} \times \int_0^1 \frac{d\lambda}{-c + u_l + \Delta u.\lambda}$$

which is the desired jump condition. In the more complicated case of the form $a(x)(H_\sigma)' + b(x)(H_\sigma) + c(x) = 0$ one has first to resolve this equation as usual. One obtains also $H_\sigma$ in function of $H_u$. <u>Therefore this shows that the simplification in section 9 (due to the fact the first 2 equations there are in conservation form) is unimportant, although welcome for simpler calculations</u>. For systems of equations in nonconservation form stated with = in **G** the calculations that involve several physical variables are more complicated but similar to the simple calculation in this section.

**13-Conclusion**. Most of this material has been extracted from industrial works on typical nonlinear problems in order to show
*to <u>mathematicians:</u> **examples of use of nonlinear generalized functions** (there is still a widespread belief among mathematicians that "there will never exist a nonlinear theory of generalized functions in any mathematical context", see[C1,G-K-O-S]);
*to <u>applied mathematicians and engineers:</u> **standard models that can be reproduced in similar situations.**
  Sections 4,5,7 (with the calculations sketched in sections 9,11,12) concern the design of armour. Section 6 concerns a mixture of oil, water and gas in pipes. Section 8 could serve as a starting model to predict the trajectory of hurricanes.
  Linear problems are not considered in this paper. Industrial applications of nonlinear generalized functions to wave propagation problems modelled by linear systems of PDEs with discontinuous coefficients are sketched in [C2§5,C1]. The coefficients represent the discontinuities of the medium. This concerns sonars, radars, earthquakes. It motivates a mathematical extension of "microlocal analysis" to the case of linear PDEs with irregular coefficients as a mathematical tool for the above engineering problems.

A large part of this research has been done inside the team of A.Y. LeRoux [LR1] to whom the A. is greatly indebted.

jf.colombeau@wanadoo.fr; 33 rue de la Noyera, pavillon 17, 38090, Villefontaine, France.

**References**.